\documentclass[longauth]{aa}  

\usepackage{linenoaa}   
\usepackage{orcidlink}
\usepackage{graphicx}   
\usepackage{amssymb}    
\usepackage{amsmath}    
\usepackage{mathrsfs}   
\usepackage{grffile} 
\usepackage{xcolor}
\usepackage{lscape}
\usepackage{placeins}
\usepackage{multirow} 
\usepackage[caption=false]{subfig}
\usepackage{hyperref}
\usepackage{txfonts}
\begin{document}

   \title{Studying baryon acoustic oscillations using photometric redshifts from the DESI Legacy Imaging survey DR9}
   \titlerunning{BAO using photometric redshifts from DESI DR9}

   \author{Christoph Saulder\orcidlink{0000-0002-0408-5633}
          \inst{1,2}
          \and
          Yong-Seon Song\inst{3}\fnmsep\thanks{Corresponding author: \email{ysong@kasi.re.kr}}
          \and
          Minji Oh\orcidlink{0000-0003-4181-138X} \inst{4}
          \and
          Yi Zheng \inst{5,6}
          \and
          Ashley J. Ross\orcidlink{0000-0002-7522-9083} \inst{7} 
          \and
          Rongpu Zhou\orcidlink{0000-0001-5381-4372} \inst{8}
          \and
          Jeffrey A. Newman\orcidlink{0000-0001-8684-2222} \inst{9}
          \and
          Chia-Hsun Chuang\orcidlink{0000-0002-3882-078X} \inst{10}      
          \and  
          Jessica Nicole Aguilar \inst{8}
          \and          
          Steven Ahlen\orcidlink{0000-0001-7145-8674} \inst{11}      
          \and
          Robert Blum\orcidlink{0000-0002-8622-4237} \inst{12}
          \and
          David Brooks \inst{13}
          \and
          Todd Claybaugh \inst{8}
          \and
          Axel de la Macorra\orcidlink{0000-0002-1769-1640} \inst{14}
          \and
          Biprateep Dey\orcidlink{0000-0002-5665-7912} \inst{9}
          \and 
          Zhejie Ding\orcidlink{0000-0002-3369-3718} \inst{15} 
          \and
          Peter Doel \inst{13}
          \and
          Jaime E. Forero-Romero\orcidlink{0000-0002-2890-3725} \inst{16,17}
          \and          
          Enrique Gaztañaga \inst{18,19,20}
          \and
          Satya Gontcho A Gontcho\orcidlink{0000-0003-3142-233X} \inst{8}
          \and
          Gaston Gutierrez \inst{21}
          \and 
          Stephanie Juneau \inst{12}
          \and
          David Kirkby\orcidlink{0000-0002-8828-5463} \inst{22}
          \and
          Theodore Kisner\orcidlink{0000-0003-3510-7134} \inst{8}
          \and
          Anthony Kremin\orcidlink{0000-0001-6356-7424} \inst{8}
          \and 
          Andrew Lambert \inst{8}
          \and
          Martin Landriau\orcidlink{0000-0003-1838-8528} \inst{8}
          \and
          Laurent Le Guillou\orcidlink{0000-0001-7178-8868} \inst{23}
          \and
          Michael Levi\orcidlink{0000-0003-1887-1018} \inst{8}
          \and
          Aaron Meisner\orcidlink{0000-0002-1125-7384} \inst{12}
          \and
          Eva-Maria Mueller \inst{24}
          \and
          Andrea Muñoz-Gutiérrez \inst{25}
          \and 
          Gustavo Niz\orcidlink{0000-0002-1544-8946} \inst{26,27}
          \and
          Francisco Prada \inst{28}
          \and
          Mehdi Rezaie\orcidlink{0000-0001-5589-7116} \inst{29}
          \and
          Graziano Rossi \inst{30}
          \and
          Eusebio Sanchez\orcidlink{0000-0002-9646-8198} \inst{31}
          \and 
          Michael Schubnell \inst{32}
          \and
          Joseph Harry Silber\orcidlink{0000-0002-3461-0320} \inst{8}
          \and
          David Sprayberry \inst{12}
          \and
          Gregory Tarl\'{e} \inst{32}
          \and
          Francisco Valdes\orcidlink{0000-0001-5567-1301} \inst{12}
          \and
          Benjamin Alan Weaver \inst{12}
          \and
          Hu Zou\orcidlink{0000-0002-6684-3997} \inst{33}
}
   \institute{Max Planck Institute for Extraterrestrial Physics, Gie\ss enbachstra\ss e 1, 85748 Garching, Germany 
         \and
             Universit\"ats-Sternwarte M\"unchen, Scheinerstra\ss e 1, 81679 Munich, Germany 
        \and
        Korea Astronomy \& Space Science Institute, 776 Daedeokdae-ro, Yuseong-gu, 34055 Daejeon, Republic of Korea 
        \and
        Chosun University, Chosundaegil 146, Dong-gu, 61452 Gwangju, Republic of Korea 
        \and
  School of Physics and Astronomy, Sun Yat-sen University, 2 Daxue Road, Tangjia, Zhuhai, 519082, China 
  \and
    CSST Science Center for the Guangdong-Hong kong-Macau Greater Bay Area, SYSU, China 
    \and
    Center for Cosmology and AstroParticle Physics, The Ohio State University, Columbus, OH 43210, USA 
    \and 
    Lawrence Berkeley National Laboratory, 1 Cyclotron Road, Berkeley, CA 94720, USA 
    \and
    Department of Physics and Astronomy and PITT PACC, University of Pittsburgh, 3941 O’Hara St., Pittsburgh, PA 15260, USA 
    \and
    Kavli Institute for Particle Astrophysics and Cosmology, Stanford University, 452 Lomita Mall, Stanford, CA 94305, USA 
    \and
    Physics Dept., Boston University, 590 Commonwealth Avenue, Boston, MA 02215, USA 
    \and
    NSF NOIRLab, 950 N. Cherry Ave., Tucson, AZ 85719, USA 
    \and 
    Department of Physics \& Astronomy, University College London, Gower Street, London, WC1E 6BT, UK 
    \and
    Instituto de F\'{\i}sica, Universidad Nacional Aut\'{o}noma de M\'{e}xico,  Cd. de M\'{e}xico  C.P. 04510,  M\'{e}xico 
    \and 
    Department of Astronomy, School of Physics and Astronomy, Shanghai Jiao Tong University, Shanghai 200240, China 
\and
Departamento de F\'isica, Universidad de los Andes, Cra. 1 No. 18A-10, Edificio Ip, CP 111711, Bogot\'a, Colombia 
\and
Observatorio Astron\'omico, Universidad de los Andes, Cra. 1 No. 18A-10, Edificio H, CP 111711 Bogot\'a, Colombia 
       \and
Institut d'Estudis Espacials de Catalunya (IEEC), 08034 Barcelona, Spain 
  \and
Institute of Cosmology and Gravitation, University of Portsmouth, Dennis Sciama Building, Portsmouth, PO1 3FX, UK 
  \and
Institute of Space Sciences, ICE-CSIC, Campus UAB, Carrer de Can Magrans s/n, 08913 Bellaterra, Barcelona, Spain 
\and
Fermi National Accelerator Laboratory, PO Box 500, Batavia, IL 60510, USA 
\and
Department of Physics and Astronomy, University of California, Irvine, 92697, USA 
\and
Sorbonne Universit\'{e}, CNRS/IN2P3, Laboratoire de Physique Nucl\'{e}aire et de Hautes Energies (LPNHE), FR-75005 Paris, France 
\and
Department of Physics and Astronomy, University of Sussex, Brighton BN1 9QH, UK 
\and
Instituto de F\'{\i}sica, Universidad Nacional Aut\'{o}noma de M\'{e}xico,  Cd. de M\'{e}xico  C.P. 04510,  M\'{e}xico 
\and
Departamento de F\'{i}sica, Universidad de Guanajuato - DCI, C.P. 37150, Leon, Guanajuato, M\'{e}xico 
\and
Instituto Avanzado de Cosmolog\'{\i}a A.~C., San Marcos 11 - Atenas 202. Magdalena Contreras, 10720. Ciudad de M\'{e}xico, M\'{e}xico 
\and
Instituto de Astrof\'{i}sica de Andaluc\'{i}a (CSIC), Glorieta de la Astronom\'{i}a, s/n, E-18008 Granada, Spain 
\and 
Department of Physics, Kansas State University, 116 Cardwell Hall, Manhattan, KS 66506, USA 
\and
Department of Physics and Astronomy, Sejong University, Seoul, 143-747, Korea 
\and
CIEMAT, Avenida Complutense 40, E-28040 Madrid, Spain 
\and
University of Michigan, Ann Arbor, MI 48109, USA 
\and
National Astronomical Observatories, Chinese Academy of Sciences, A20 Datun Rd., Chaoyang District, Beijing, 100012, P.R. China 
}

   \date{Received 27/08/2024; accepted 17/01/2025}

 
  \abstract
  {The Dark Energy Spectroscopic Instrument (DESI) Legacy Imaging Survey DR9 (DR9 hereafter), with its extensive dataset of galaxy locations and photometric redshifts, presents an opportunity to study baryon acoustic oscillations (BAOs) in the region covered by the ongoing spectroscopic survey with DESI.} 
   {We aim to investigate differences between different parts of the DR9 footprint. Furthermore, we want to measure the BAO scale for luminous red galaxies within them. Our selected redshift range of 0.6 to 0.8 corresponds to the bin in which a tension between DESI Y1 and eBOSS was found.}
   {We calculated the anisotropic two-point correlation function in a modified binning scheme to detect the BAOs in DR9 data. We then used template fits based on simulations to measure the BAO scale in the imaging data.}
   {Our analysis reveals the expected correlation function shape in most of the footprint areas, showing a BAO scale consistent with \textit{Planck}'s observations. Aside from identified mask-related data issues in the southern region of the South Galactic Cap, we find a notable variance between the different footprints.}
   {We find that this variance is consistent with the difference between the DESI Y1 and eBOSS data, and it supports the argument that that tension is caused by sample variance. Additionally, we also uncovered systematic biases not previously accounted for in photometric BAO studies. We emphasize the necessity of adjusting for the systematic shift in the BAO scale associated with typical photometric redshift uncertainties to ensure accurate measurements.}


   \maketitle
%
\nolinenumbers

\section{Introduction} \label{sec:intro}

Measuring the expansion history of the Universe is of paramount importance in the field of modern cosmology. It will allow us to improve our understanding of the nature of dark energy and gravity itself. It can be revealed by diverse cosmic distance measures in tomographic redshift space, such as cosmic parallax~\citep{Parallax}, standard candles~\citep{Standard_candle}, and standard rulers~\citep{Eisenstein_1998,Eisenstein_2005}. The best constraints to date come from the distance--redshift relation and imply that the expansion rate has changed from a decelerating phase to an accelerated one \citep{Riess_1998,Perlmutter_1999}. While current observations largely affirm the $\Lambda$-cold dark matter (CDM) model, which incorporates the cosmological constant, the quest for a high-precision confirmation of this model, or the identification of any deviations, remains a pivotal observational goal. One of the most robust methods for measuring the distance--redshift relation is to use the baryon acoustic oscillation (BAO) feature that is observed as a bump in the two-point correlation function or as wiggles in the power spectrum. The tension between gravitational infall and radiative pressure caused by the baryon-photon fluid in the early Universe gave rise to an acoustic peak structure that was imprinted on the last-scattering surface \citep{Peebles_Yu}. Acoustic features on density fluctuations are formed by a tension between the gravitational force from dark matter clustering and the radiation pressure from baryon-radiation plasma during the radiation-dominated epoch, and this length scale of the sound horizon, $r_d$, is frozen as a standard ruler at the decoupling epoch to remain in the large-scale structure of the Universe. Notably, this scale manifests as a distinct peak in the isotropic two-point correlation function. However, by analyzing the anisotropic wedge correlation function, we can discern the perpendicular and parallel components of cosmic distance. The BAO peak positions convert to cosmic distances of $D_{A}(z)/r_{d}$ and $H^{-1}(z)r_{d}$ in perpendicular and parallel directions. The quoted $r_d$ in this manuscript is fixed to $147.21$ Mpc, as obtained from \citet[hereafter Planck 2015]{Planck:2015}. The BAO feature has been measured through the correlation function \citep{Eisenstein_2005}, and one of the most successful measurements in the clustering of large-scale structure at low redshifts was obtained using data from  Sloan Digital Sky Survey (SDSS; \citealt{Eisenstein_2005,Estrada_2009,Padmanabhan_2012,Hong_2012,Veropalumbo_2014,Veropalumbo_2016}). The final results from the SDSS-III Baryon Oscillation Spectroscopic Survey (BOSS; \citealt{Ross:2014,Alam_2017}) and the SDSS-IV Extended Baryon Oscillation Spectroscopic Survey (eBOSS; \citealt{Bautista:2021,GilMarin:2020,Raichoor:2021,deMattia:2021,Hou:2021,Neveux:2020,duMasdesBourboux:2020}) traced the expansion of the Universe and showed an overall agreement with the $\Lambda$-CDM model. The Dark Energy Spectroscopic Instrument (DESI; \citealt{Levi:2013,DESI,DESI_instrument0,DESI_instrument,Silber:2023,Miller:2023}) is the main instrument of an ongoing survey \citep{Guy:2023,Schlafly:2023}. The first early data of this survey have already been released to the public consisting of the data collected during its survey validation \citep{DESI_EDR_1,DESI_EDR_2}. Furthermore, the first cosmological results based on BAO measurements \citep{DESI_2024_2,DESI_2024_3,DESI_2024_4,DESI_2024_5,DESI_2024_6,DESI_2024_7} of the first year of main survey operations have also been published and provide unprecedented precision; the full data release is still in preparation (DESI Collaboration et al, in preparation). While the results are still compatible with the $\Lambda$-CDM model, they show, when combined with other probes, a preference for a $w_{0}w_{a}$-CDM cosmology \citep{ChevallierPolarski:2001,Linder:2003,dePutter:2008}. Furthermore, there is a tension between the DESI and eBOSS BAO data around an effective redshift of 0.7. 

While the first year DESI data (DESI Y1) only cover a fraction of its final footprint, the full DESI photometric footprint has already been covered by the Legacy Imaging Surveys (\citet{DESI_Imaging,LS_DR9}, Schlegel et al., in preparation). Photometric surveys provide more observed galaxies compared to spectroscopic surveys, even at deeper redshifts \citep{Euclid_Srivatsan_2019}, but the uncertainty on the redshift obtained from photometric surveys is larger than the uncertainty on the redshift obtained from spectroscopic surveys. These photometric redshifts are measured with a much poorer resolution, which causes unpredictable damping of clustering at small scales and a smearing of the BAO peak \citep{Estrada_2009}. However, possible BAO signatures that the photometric redshift uncertainty has not washed out might still be present in the data. 

Thanks to recent advancements, we improved upon the methodology and formulation that was applied in \cite{Srivatsan:2019} to simulated photometric galaxy catalogs to get cosmological distance constraints. \citet{Chan:2022} show that using the anisotropic correlation function expressed in terms of the projected distance significantly improves the consistency of the data between different wedges. For this study, this allowed us to make improvements not only in terms of data but also in terms of methods compared to \citet{Sridhar:2020}. 

We first discuss the observational and simulated data used for our work in Sect. \ref{sec:data}. We then introduce the methods used in Sect. \ref{sec:methods} and highlight the little-discussed systematic shift of the BAOs for photometric redshift data in Sect. \ref{sec:peak_shift}. We then present our results in Sect. \ref{sec:results} and discuss them in detail in Sect. \ref{sec:discussion}. We provide a brief summary and conclusions in Sect. \ref{sec:summary}. Furthermore, we present some additional figures in the Appendix \ref{sec:appendix1}. 

\section{Observed and mock catalogs} \label{sec:data}

The DESI Legacy Imaging Survey Data Release 9 (DR9 hereafter) data are introduced in this section along with the simulations that are used to support data analysis, such as the {\tt EZmocks} and the {\tt AbacusSummit} simulation data as well as some additional dark matter simulations specifically created to test our methods.

\subsection{Photometric catalogs}

The DR9 (\citealt{DESI_Imaging,LS_DR9}) provides imaging data for the entire DESI footprint and beyond by including additional areas used by the Dark Energy Survey (DES; \citealt{DES}). Because of the large sky coverage of DR9, it was observed using three different telescopes:
\begin{itemize}
    \item The Bok 2.3 m telescope on Kitt Peak was used for the Beijing-Arizona Sky Survey (BASS; \citealt{BASS}), which observed at the sky on the North Galactic Cap (NGC) for declinations $\geq$ 32.375$^{\circ}$ using optical bands g and r to a depth of 24.0 mag and 23.4 mag, respectively. 
    \item The 4-meter Mayall telescope, which is also on Kitt Peak, was used to conduct the Mayall z-band Legacy Survey (MzLS) for the same footprint as Bok but using the z band up to a depth of 22.5 mag. 
    \item For the majority of the observations the Blanco 4m telescope at the Cerro Tololo Inter-American Observatory was used. This is the same instrument that also carried out the observations for DES \citep{DES}, but extended the footprint to incorporate areas of the NGC at declinations $\leq$ 32.375$^{\circ}$ and also areas of the South Galactic Cap (SGC) at declinations$\leq$ 34$^{\circ}$, to create the Dark Energy Camera Legacy Survey (DECaLS). The depths of these observations were 24.0 mag, 23.4 mag, and 22.5 mag in the g, r, and z bands, respectively. 
\end{itemize}
In addition to the optical imaging, near-infrared-bands W1 and W2 were added using four years of Wide-Field Infrared Survey Explorer (WISE, \citet{WISE,Wise_meisner}) data, which contributed to determine more precise photometric redshifts and target selection. In addition to DR9, we used the photometric redshifts based on DR9 data provided by \citet{Zhou:2021}, which were obtained using a random forest-based methodology \citep{Breiman:2001}.

\subsection{Simulations}
We used several different simulations to test and calibrate our methods. {\tt AbacusSummit} \citep{Abacus_sims}, a comprehensive cosmological N-body simulation, is designed to support large-scale surveys like DESI. Using the Abacus N-body code \citep{Abacus_code1,Abacus_code2,Abacus_code3} as outlined in \citealt{Abacus_first}, this suite consists of 150 box simulations across various cosmologies. For our study, we utilized 25 simulations based on their base cosmology, informed by the \citet{Planck_2018} results. These simulations feature $6912^{3}$ particles in boxes measuring 2 Gpc/h on each side. Specifically, we focused on the simulation run at a redshift of 0.8. Based on this run, mock catalogs implemented the halo occupation distribution (HOD) model (\citealt{Abacus_HOD}) for DESI luminous red galaxies (LRGs) and adapted to a cut-sky geometry, to closely mirror observational conditions.

To obtain covariance matrices for our BAO fits, we used 1000 {\tt EZmocks} \citep{EZmocks}. They are relatively accurate mock catalogs that can be generated quickly using the Zel'dovich approximation \citep{Zeldovich:1970}. We used {\tt EZmocks} that are similar to {\tt AbacusSummit} N-body simulations. 

For the initial tests of our method, we created a set of 100 simulations, called {\tt G2P}\footnote{\href{https://cosmology.kasi.re.kr/sim/sim.php}{https://cosmology.kasi.re.kr/sim/sim.php}}. We used {\tt Gadget2} by \citet{Springel:2005}, which is a code for cosmological N-body simulations. It is fed by initial conditions generated based on Lagrangian perturbation theory up to second order using {\tt 2LPTic} by \citet{Crocce:2006}, given an input power spectrum from {\tt CAMB} by \citet{Lewis:1999bs}. For the initial conditions, $N_{p}=1024^3$ number of particles are populated at $z=49$ in the simulations box of one side, 1.89Gpc/h with $N_{\textrm{mesh}}=1024$ for the number of grids in the fast Fourier Transform. 1.89Gpc/h is corresponding to the shell volume at $z=0.9$ with the redshift bin size $\Delta z = 0.2$, where the DESI survey is expected to have the largest target number density in Table 3 of \citet{Levi:2013}. After implementing the initial condition, the particles evolve with a mesh size of 2048 for {\tt TreePM} \citep{TreePM} and a softening length of 92.28 kpc, which is 5\% of the mean particle displacement, in a periodic boundary condition. We created 100 realizations using a cosmology based on \citet[Table 3]{Planck:2015} with the following parameters: $\Omega_{M}=0.3132$, $\Omega_{\Lambda}=0.6868$, $\Omega_{b}=0.049$, and H$_{0}=67.31$ (km/s)/Mpc. 

To generate the halo catalog, we applied the group finder {\tt ROCKSTAR}~\cite{rockstar} to each simulation run. We found all halos with at least ten dark matter particles. The halo position is evaluated by averaging the particle locations for the inner subgroup that best minimizes the Poisson error. The halo velocity is calculated by using the average particle velocity within the innermost $10\%$ of the halo radius. To fill these dark matter only simulations with a realistic representation of galaxies, we applied a HOD (\citealt{Berlind:2002}) model sample of LRGs that will be observed by DESI. For this model, we used the HOD parameters from \citet{Zhou:2021} for a photometric redshift range between 0.61 and 0.72 and applied them to our simulations using the {\sc halotools} implementation of the HOD model from \citet{Zheng:2007}.

\section{Methods}
\label{sec:methods}
\subsection{Sample selection of the photometric data}
\label{sec:sampleselection}
We selected a subset of the DESI LRG sample. We followed the official target selection outlined in \citet{DESI_target_LRG}, which consists of a fiber magnitude cut, a series of color cuts (to remove stars and select red galaxies in the right redshift range), and several quality checks (photometric information in all bands and masking of foreground objects). In addition to these selection criteria, we also applied masks for objects near bright stars \footnote{MASKBITS 8 and 11 according to the bit masks of the DESI Legacy Imaging Survey
\href{https://www.legacysurvey.org/dr9/bitmasks/}{https://www.legacysurvey.org/dr9/bitmasks/}}. We decided to limit the range of sample to photometric redshifts \citep{Zhou:2021} between 0.6 and 0.8 and thereby obtain an effective redshift of 0.701. 

\begin{figure}
    \centering
        \includegraphics[width=0.48\textwidth]{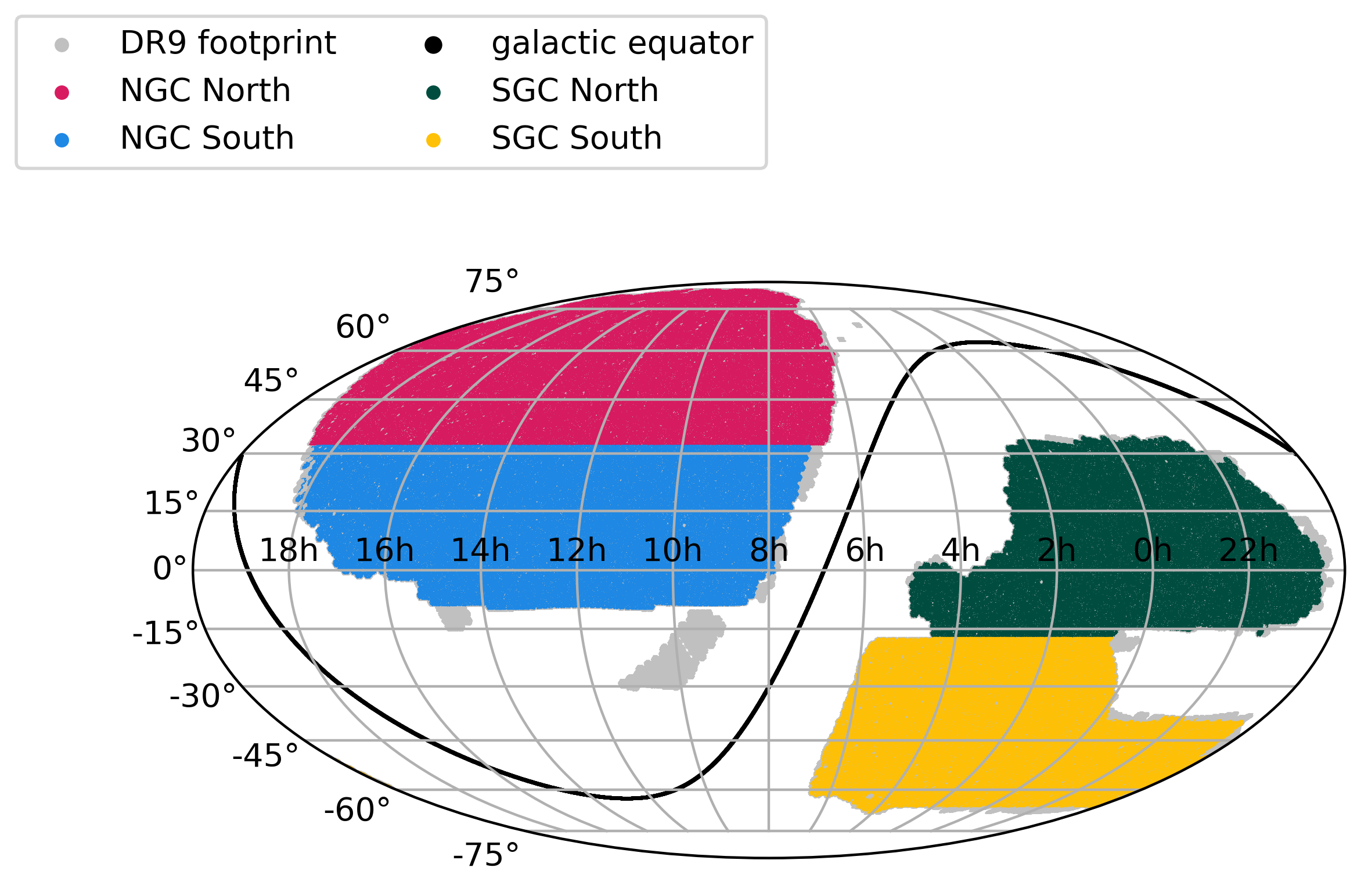}  
        \caption{Sky map of our different footprint subsamples. The NGC North footprint corresponds to the area covered by the Bok and MzLS surveys, while all other footprints were observed as part of DECaLS. The split between SGC North and SGC South is motivated by the planned reach of DESI spectroscopy.}
        \label{fig:skymap}   
\end{figure} 

\begin{table}
\begin{center}
\caption{Comparison of the different footprint subsamples.}
\label{tab:samplesize}
\begin{tabular}{cccc} 
sample name & survey source & sky area & galaxies \\
\hline
NGC North & BASS/MzLS & $\sim4675$ deg$^{2}$ & 889 672\\ 
NGC South & DECaLS & $\sim5344$ deg$^{2}$ & 1 016 011\\ 
SGC North & DECaLS & $\sim4358$ deg$^{2}$ & 824 222\\ 
SGC South & DECaLS & $\sim3863$ deg$^{2}$ & 703 239\\ 
\end{tabular}
\tablefoot{They are of roughly comparable size.}
\end{center}
\end{table}

In the next step, we divided the dataset into four subsamples of comparable size (see Table \ref{tab:samplesize}). The data from the BASS/MzLS with declinations greater than 32.375 deg, to avoid overlap, form the NGC North sample. As they were collected by different telescopes, these data have distinct photometric characteristics that required minor target selection adjustments  (see \citealt{DESI_target_LRG}) and have a different error budget for the photometric redshifts. We further divided the data collected from DECaLS into one sample on the NGC and two samples on the SGC. The NGC South sample consists of the DECaLS data on the NGC between a declination of 32.375 deg (to avoid overlap with NGC North) and -10 deg (to avoid a few disconnected patches farther south). We split the data for the SGC along a declination of -17 deg into the SGC North and SGC South areas. The SGC North roughly corresponds to the area of the SGC that will be covered by the spectroscopic DESI survey. We provide an overview map of these samples in Fig. \ref{fig:skymap}.

\subsection{The treatment of the random catalogs}
\begin{figure*}
    \centering
        \includegraphics[width=0.95\textwidth]{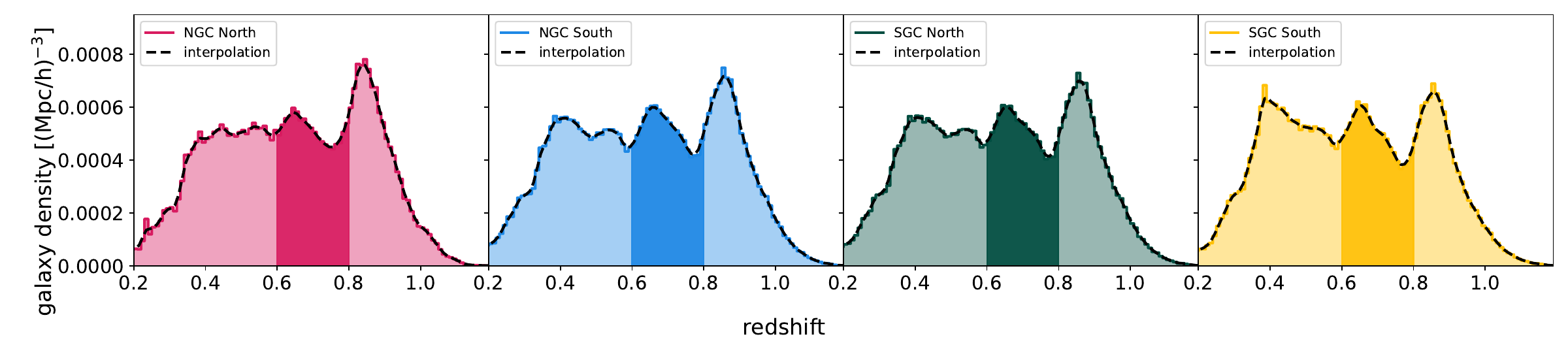}  
        \caption{Number density of LRGs as a function of photometric redshifts. First (leftmost) panel: Distribution for BASS/MzLS with the redshift slice used in our analysis highlighted. Dotted black lines show the smoothed interpolation function used for populating the redshift distribution in the random catalogs. Second panel: Same but for DECaLS within the NGC. Third panel: Same but for DECaLS within the northern part of the SGC. Forth panel: Same but for DECaLS within the southern part of the SGC.}
        \label{fig:density}   
\end{figure*} 

As we aimed to calculate correlation functions for these data, we needed suitable random catalogs. The DESI Legacy Imaging Survey provides 20 random catalogs with a density of 2 500 points per square degree covering its entire footprint. Additionally, these random catalogs contain the same mask and photometric pass information as the data. Hence, we simply applied the same selection for the mask bits and pass information as we did for the observational data. To match our four samples of different regions, we applied the same geometric cuts to the randoms as well. Furthermore, we used the number counts of the real data to define the density of each of the randoms in (photometric) redshift space. As the uncertainties and densities are slightly different based on the telescopes used to collect the imaging data, we treated the BASS/MzLS (essentially NGC North) and DECaLS (all the others) areas separately. We used the observed densities, applied a simple Gaussian smoothing, and used cubic splines for interpolation between the bin centers (see Fig. \ref{fig:density}). We used this smoothed distribution to populate them with the suitable redshift distribution and adjusted the density accordingly. In the last step, we combined the 20 random catalogs of each footprint into one big random catalog for each footprint.

\subsection{Sample selection in simulations}
We focused on the cut-sky simulations that were generated from the z=0.8 snapshot with a LRG HOD applied to them. They were created specifically for the DESI to validate its pipeline on mock catalogs. As the photometric footprint is further extended than the spectroscopic footprint of DESI, we had to create suitable masks for it in the {\tt EZmocks} and {\tt AbacusSummit} simulations. To this end, we used the randoms from the DESI Legacy Imaging survey to create a {\tt HEALPix} map \citep{HEALPix} with {\tt nside} of 256 to determine which parts of the sky are covered by our samples. We selected all {\tt HEALPix} cells that have at least one random point in them to be the mask of the photometric footprint. An additional advantage that this method has is that it also retains the biggest hole created by the various {\tt MASKBITS} (typically large galaxies and bright stars) as well and not just the overall shape of the distribution. Additional geometric cuts to recreate our four distinct samples were carried out after the mask was applied to the cut-sky simulations. In addition to the projected distribution on the sky, we also wanted to recover the redshift-space distribution correctly, and therefore, we need to understand the scatter created by the large uncertainties of the photometric redshifts. 

\subsection{Photometric redshift painting}
\label{sec:photozpaint}
\begin{figure*}
    \centering
        \includegraphics[width=0.95\textwidth]{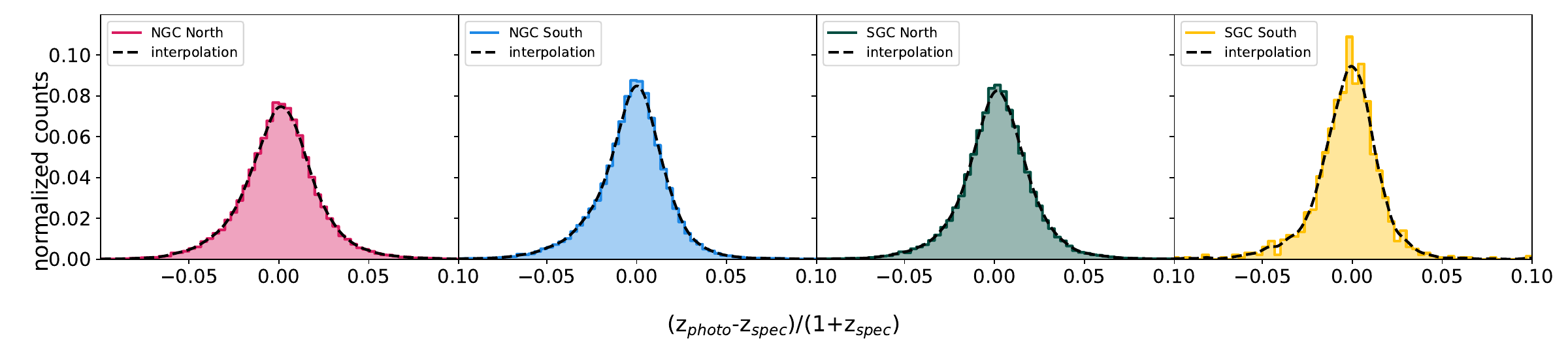}  
        \caption{Uncertainties of the photometric redshifts for the 0.6 to 0.8 redshift range. First (leftmost) panel: Distribution for BASS/MzLS. Dotted black lines show the smoothed interpolation function used for the photometric redshift painting. Second panel: Same but for DECaLS within the NGC. Third panel: Same but for DECaLS within the northern part of the SGC. Forth panel: Same but for DECaLS within the southern part of the SGC.}
        \label{fig:photozpaint}   
\end{figure*} 

As the photometric redshifts are the dominant effect that changes the shape of the correlation function, we needed to reproduce the scatter created by them in the redshift distribution as well as reasonably possible in the simulations. We analyzed the photometric redshifts, $z_{\textrm{photo}}$, within our LRG sample and our selected redshift range and compared them to the spectroscopic redshifts $z_{\textrm{spec}}$ of the training sample used by \citet{Zhou:2021} to calibrate them. The distribution of $(z_{\textrm{photo}}-z_{\textrm{spec}})/(1+z_{\textrm{spec}})$, which are shown, for both the BASS/MzLS and the DECaLS separately, in Fig. \ref{fig:photozpaint} can be used to apply the correct scatter to the redshifts in the simulations. The distribution roughly corresponds to a Gaussian with a width $\sigma_{0}$ of about 0.02. However, as there are deviations in its Gaussian shape, we decided to use a smoothed interpolation of the measured distribution instead of a Gaussian interpolation to reproduce the scatter created by the uncertainty of the photometric redshift estimates. We drew the photometric scatter $\Delta z$ for the interpolated distribution and add it to the redshift $z_{\textrm{rsd}}$ obtained from the simulations using 
\begin{equation}
(1+z_{\textrm{photo}})=(1+\Delta z)\cdot(1+z_{\textrm{rsd}}).
\label{eq:photozadd}
\end{equation}
The $z_{\textrm{rsd}}$ is already taken into account the redshift space distortion introduced by the peculiar motions of the galaxies. We applied this to all our datasets from the {\tt AbacusSummit} cut-sky simulation and {\tt EZmocks}. In the case of the box simulations, which we used only for complementary tests of our methods, we went for the simplified method using a simple Gaussian distribution with $\sigma_{0}=0.02$ instead and applied it to the z-direction taking into account the periodic boundary conditions and scaling of the box given its cosmology. Based on the overall characteristics of the photometric redshifts \citep{Zhou:2021}, we assumed the redshift evolution of the scatter of the photometric redshifts slice to be sufficiently small within our ultimate redshift.

\subsection{Correlation function measurements}
Two point correlation function $\xi(\textbf{r})$ \citep{totsuji_1969,davis_peebles_1983} measures the excess probability of two galaxies separated by a cosmic distance \textbf{r}, relative to the Poisson expectation, which exhibits an anisotropic distortion along the line of sight (LOS) at redshift space. The two-point correlation function is typically estimated using the Landy \& Szalay estimator (hereafter LS; \citealt{landy_szalay}): 
\begin{equation}
\xi_{LS} = \frac{DD - 2 DR + RR}{RR}
\label{eq:landyszalay_estimator}
,\end{equation}
where $DD$ and $RR$ are the normalized number of pairs in the data and randoms catalog, respectively within a given binning scheme. $DR$ correspond to pairs between the data ($D$) and random catalog ($R$). Throughout this work, we used random catalogs of a density 20 times the data. We used the Planck 2015 cosmology for the redshift distance relation:
\begin{align}
s^2=s_{\perp}^2+s_{\parallel}^2 \nonumber \\
\mu=s_{\parallel}/s. 
\label{eq:mu_definiton}
\end{align}
In this study we used the anisotropic correlation function, which is typically binned in $(s,\mu)$ coordinates, where $s$ denotes the radius to the shell and $\mu$ denotes the observed cosine of the angle to the galaxy pair. The transverse ($s_{\perp}$) and radial ($s_{\parallel}$) separations are related to these coordinates by Eq. \ref{eq:mu_definiton}. While spectroscopic data with minimal scatter along the LOS enables a consistent recovery of BAO features in the $\xi(s,\mu)$ function, the increased scatter in photometric redshift measurements tends to wash out these features, especially along the LOS. As demonstrated in \citet{Srivatsan:2019} and \citet{Sridhar:2020}, sufficient information perpendicular to the LOS remains to discern the BAO peak. However, \citet{Chan:2022} show that in the case of data with large scatter, the correlation function between the various $\mu$ bins is more consistent in $(s_{\perp},\mu)$ coordinates than $(s,\mu),$ with $s_{\perp}$ being the transverse separation (perpendicular to the LOS). As our own tests with simulated data confirm this, we consequently calculated all correlation functions that use photometric redshifts as functions of $(s_{\perp},\mu)$, while for data without large uncertainties (such as simulated spectroscopic data) we calculated the correlation function in $(s,\mu)$ coordinates. 

Throughout this work, unless stated otherwise, we used the following binning scheme: in $s$ or $s_{\perp}$ we used 13 equal-sized bins of 7 Mpc/h width ranging between 42.5 and 133.5 Mpc/h, with bin centers at 46, 53, ... 130 Mpc/h. We used five equal-size $\mu$ bins between 0 and 0.5 with bin centers at $\bar{\mu}$=0.05, 0.15, ... 0.35; using too many $\mu$ bins would complicate the covariance matrix, and using too few $\mu$ bins would not allow us to separate the error propagation along the LOS clearly \citep{Cris_2016}. The higher $\mu$ between 0.5 and 1 are not used as there is too little information retained in them. The large photo-z error destroys the radial BAO signal efficiently, while still retaining some angular BAO signal. Hence, with increasing $\mu$, as we are probing more and more of the radial component, the pair counts decrease for the same separation. The calculations of the correlation functions were performed with the Python package \emph{corrfunc}\footnote{\url{https://github.com/manodeep/Corrfunc}} \citep{corrfunc,corrfunc2}, as well as additional code, creates a wrapper around the $\xi(s_{\perp},s_{\parallel})$\footnote{In the literature, such as the \emph{corrfunc} documentation, $s_{\perp}$ is called $\sigma$ and $s_{\parallel}$ is called $\pi$.} calculations of \emph{corrfunc} to obtain $\xi(s_{\perp},\mu)$.

\subsection{Covariance matrix}
For the subsequent fits to the observational data (and simulations) suitable covariance matrices are required. To this end, we calculated the correlation functions for the 1000 {\tt EZmocks} for each of the four subsamples as well as using the 100 {\tt G2P} simulations for supplementary tests in Sect. \ref{sec:peak_shift}. Consequently, we calculated the covariance matrix as\begin{equation}
C^{ij}(\xi^i,\xi^j)=\frac{\sum^{N_{\textrm{mocks}}}_{n=1}[\xi^n(\vec{x}_i)-\overline{\xi}(\vec{x}_i)][\xi^n(\vec{x}_j)-\overline{\xi}(\vec{x}_j)]}{N_{\textrm{mocks}}-1},\label{eqn:cov_matrix}
\end{equation}
where the total number of simulations is given by $N_{\textrm{mocks}}$. The $\xi^n(\vec{x}_i)$ represents the value of the of the correlation function of $i^{th}$ bin of $\vec{x}_i$ in the $n^{th}$ realization and $\overline{\xi}(\vec{x}_i)$ is the mean of all the realization. $\vec{x}_i$ stands for each of the 65 combinations of the ($s$,$\mu$) or ($s_{\perp}$,$\mu$) bins, depending on which type of data is used. 

Additionally, we also accounted for the offset caused by the finite number of realization \citep{Hartlap_2007} as
\begin{equation}
C^{-1}=\frac{N_{\textrm{mocks}} - N_{\textrm{bins}}-2}{N_{\textrm{mocks}}-1}\ \hat{C}^{-1}\ ,\label{eqn:percival}
\end{equation}
where $N_{bins}$ denotes the total number of $i$ bins, which are 65 in our case. $\hat{C}^{-1}$ and $C^{-1}$ are the inverted covariance matrix before and after the correction was applied. 

Apart from the above correction factor, an additional correction to the inverse covariance matrix is proposed by \cite{Percival_2014}: 

\begin{align}
C^{-1} &=p_{\textrm{cor}} \hat{C}^{-1}\ , \nonumber\\
p_{\textrm{cor}} &= \sqrt{\frac{1+B (N_{\textrm{bins}} - N_{\textrm{para}})}{1+A+B (N_{\textrm{para}} + 1)}},\\
A&=\frac{2}{(N_{\textrm{mocks}}-N_{\textrm{bins}}-1)(N_{\textrm{mocks}}-N_{\textrm{bins}}-4)},\nonumber\\
B&=\frac{N_{\textrm{mocks}}-N_{\textrm{bins}}-2}{(N_{\textrm{mocks}}-N_{\textrm{bins}}-1)(N_{\textrm{mocks}}-N_{\textrm{bins}}-4)},\nonumber\label{eqn:percival}
\end{align}
with $N_{\textrm{para}}$ being the number of free parameters used in the fits. 

\subsection{Template fits}
\label{sec:template_fits}
\begin{figure}
    \centering
        \includegraphics[width=0.47\textwidth]{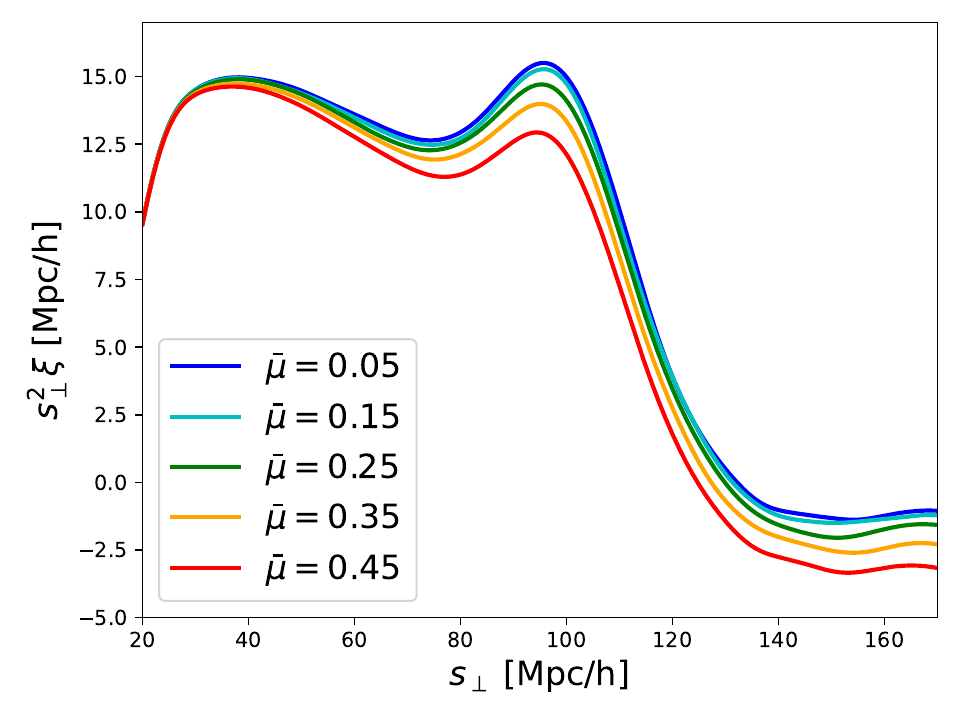}  
        \caption{Basic templates for different $\mu$ bins derived from simulations assuming a photometric redshift uncertainty of $\sigma_{0}$. They show the two-point correlation functions in terms of $s_{\perp}$ for different $\mu$ bins.}
        \label{fig:template}   
\end{figure} 
In a different approach to the previously used method \citep{Sanchez_empirical,Sanchez_empirical2,Veropalumbo_2016,Srivatsan:2019,Sridhar:2020} of fitting an empirical function to the data, we implemented a simplified template fitter that takes advantage of the whole correlation function (within our bin-range). This approach is similar to the one used in \cite{Xu:2012} and \cite{Ross:2017} but is based on simulations that take into account the photometric redshift effects. To create the templates, we used the mean of the 100 {\tt G2P} simulations with the photometric redshift painting applied to them and calculated $\xi(s_{\perp},\mu)$ at a resolution of 1 Mpc/h in $s_{\perp}$ over a much-extended range between 20 and 170 Mpc/h and for $\mu$ bins of the 0.1 width. From these simulations, we obtained interpolation functions for each $\mu$ bin, as shown in Fig. \ref{fig:template}. Following \citet{Moon:2023}, but with only one broadband term, 
\begin{equation}
\xi_{\textrm{fit}}(s_{\perp},\mu)= B \cdot \xi_{\textrm{template}}(s_{\perp} \alpha_{\perp},\mu) + A_{0} ,
\label{eq:template} 
\end{equation} 
we allowed this function to be stretched by $\alpha_{\perp}$ and shifted using the nuisance parameters $B$ and $A_{0}$. $\alpha_{\perp}$ in this case, is relative to the cosmology used to generate the simulations for the templates. For our application, we used a prior range of 0.8 to 1.2 on $\alpha_{\perp}$ and limited $B$ to positive values. All other nuisance parameters were left unconstrained. 

\subsection{Joint fits of different $\mu$ bins}
\label{sec:jointfits}
When fitting the different $\mu$ bins of $\xi(s_{\perp},\mu)$, we noticed that that data are still affected by cosmic variance. \citet{Chan:2022} show that for the level of photometric redshift uncertainty expected for the DESI Legacy Imaging Survey, we should expect all BAO peaks of different $\mu$ bins to align, if calculated as a function of $s_{\perp}$ and thereby yield the same $\alpha$-parameter. Therefore, we jointly fit the different $\mu$ bins with $\alpha_{\perp}$ being the only common parameter. We adopted a standard likelihood, $\mathscr{L} \propto \mathrm{exp}(-\chi^{2}/2)$ where the function $\chi^{2}$ is defined as
\begin{align}
\chi^{2}=\sum_{s_{\perp},s_{\perp}',\mu,\mu'} \Delta \xi (s_{\perp},\mu) C^{-1}({s_{\perp};s_{\perp}';\mu;\mu'})\Delta \xi (s_{\perp}',\mu') \\
\Delta \xi (s_{\perp},\mu) = \xi_{\textrm{mod}}(s_{\perp},\mu)-\xi_{\textrm{data}}(s_{\perp},\mu). \nonumber
\label{eq:jointlike}
\end{align}
This means that instead of several independent five-parameter fits for each $\mu$ bin, we have one $1+2 \cdot n_{\mu}$ parameter fit for $n_{\mu}$ jointly fitted $\mu$ bins since we had used the same $\alpha_{\perp}$ parameters for the joint fits to all the $\mu$ bins. 

\section{BAO peak shift}
\label{sec:peak_shift}
\begin{figure*}
    \centering
        \includegraphics[width=0.95\textwidth]{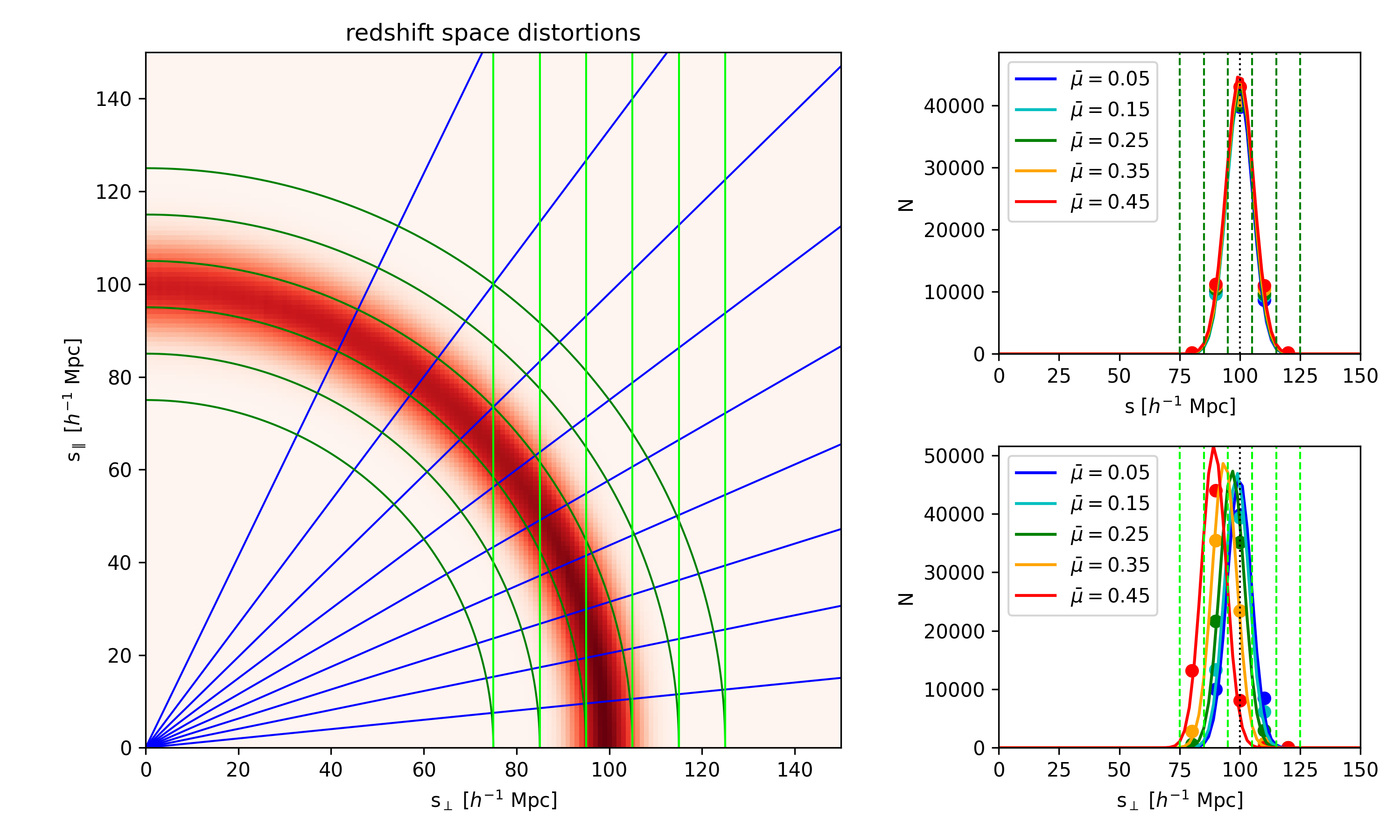}  
        \caption{Toy model for illustrations showing the anisotropic correlation function for spectroscopic redshifts. Left panel: Illustration of the BAOs after redshift space distortions are applied. Blue lines separate the $\mu$ bins. Dark green lines separate the $s$ bins. Light green lines separate the $s_{\perp}$ bins. Top-right panel: Counts within the $s,\mu$ bins. Bottom-right panel: Counts within the $s_{\perp},\mu$ bins.}
        \label{fig:toymodel_rsd}   
\end{figure*} 
\begin{figure*}
    \centering
        \includegraphics[width=0.95\textwidth]{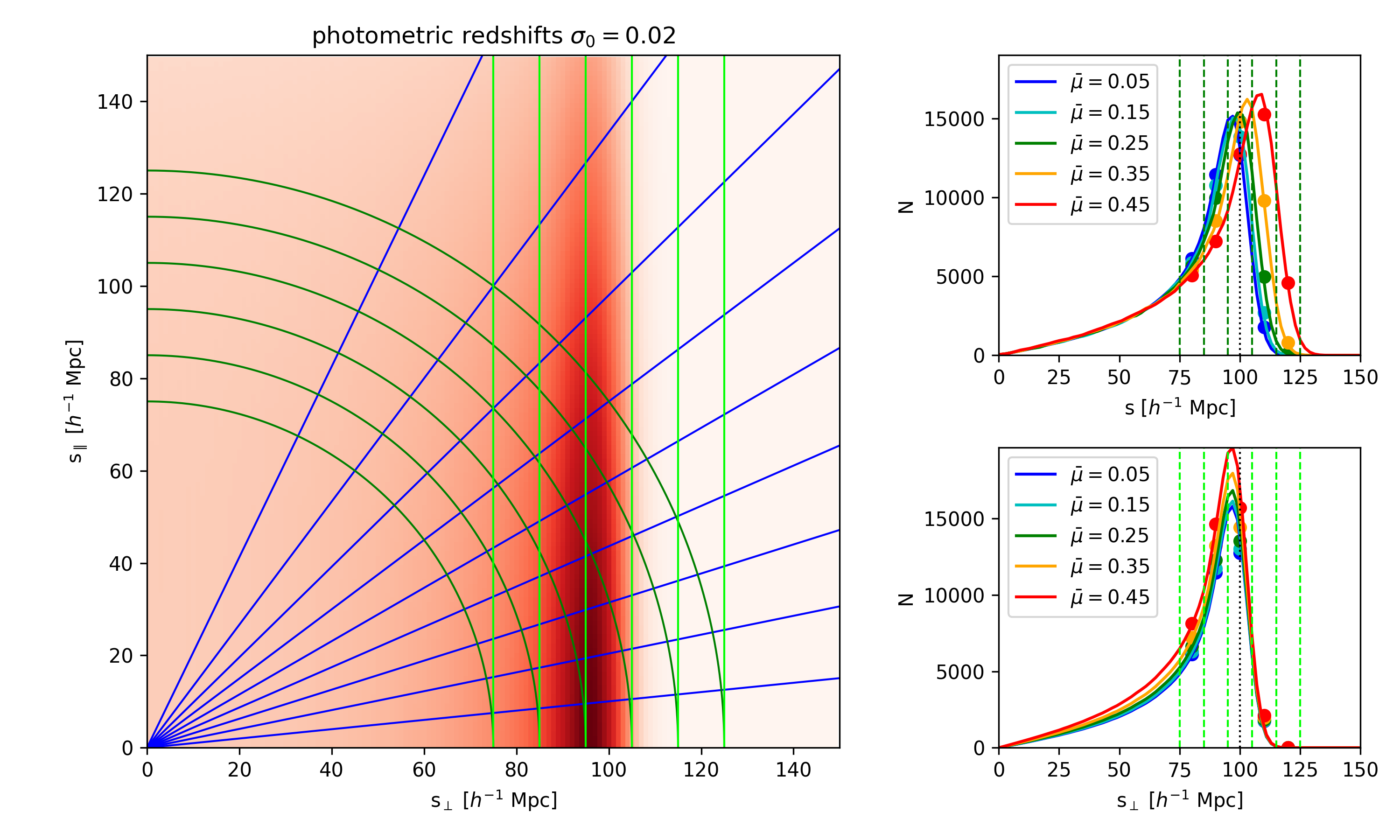}  
        \caption{Toy model for illustrations showing the anisotropic correlation function for photometric redshifts. Same as Fig. \ref{fig:toymodel_rsd} but for the BAOs after applying a photometric redshift uncertainty of $\sigma_{0}$=0.02. }
        \label{fig:toymodel_photo}   
\end{figure*} 

One of the reasons why we moved to template-based fits instead of an empirical function, was that we discovered that the location of BAO peak for data using photometric redshifts with their typical uncertainty is systematically shifted compared to the location of the BAO peak obtained from data with spectroscopic accuracy. This is the case when measuring the anisotropic two-point correlation function of the photometric data in terms of $\xi(s_{\perp},\mu)$. While \citet{Chan:2022} have already shown that the location of the BAO peak is consistent between the different $\mu$ bins, if expressed in terms of $\xi(s_{\perp},\mu)$, they did not report that the absolute location of it is shifted relative to the location of the BAO peak found using the same data, but with only redshift-space distortions (RSDs) or typical spectroscopic uncertainty accounted for. We found this effect to be present for all data within a redshift uncertainty $\sigma_{0}$ of more than $\sim 0.01$ (at the redshifts of our simulations of $0.7$). The shift is also consistent between different $\mu$ bins and between different simulations that contain enough data to get beyond the cosmic variance threshold. 

To better understand this phenomenon, we created a simple toy model that lets us explore how the larger uncertainties of the photometric redshifts affect the BAO feature. We used a $s_{\perp}$-$s_{\parallel}$ diagram, with $s_{\parallel}$ being the separation parallel to the LOS  and $s_{\perp}$ being the separation orthogonal to the LOS, as the main tool in our toy model. We then populated this diagram with only a ring-like over-density with a Gaussian profile centered around a separation of 100 (in arbitrary units) and a width of 5 (again in arbitrary units). When further applied a redshift uncertainty $\sigma_{0}=0.001$ along the LOS, which approximately corresponds to the magnitude of redshift space distortions. In the resulting map, which is shown in Fig. \ref{fig:toymodel_rsd}, we mark the different wedge-like $\mu$ bins used for the anisotropic correlation function as straight lines spreading from the origin of the coordinate system. Additionally, we mark circular bins representing the binning scheme of $\xi(s,\mu)$ and straight bins orthogonal to the LOS representing the binning scheme of $\xi(s_{\perp},\mu)$. Furthermore, Fig. \ref{fig:toymodel_rsd} also shows the number counts within the bins, representing a non-normalized correlation function, in plots on the side for both binning schemes $(s,\mu)$ and $(s_{\perp},\mu)$. For practical reasons, we only show the first 5 $\mu$ bins. We see that our toy model, which represents the BAOs detected using spectroscopic redshifts, yields consistent locations for the center of the BAO peak in the $(s,\mu)$ bins and that it is located where we set it, at 100. However, for the $\xi(s_{\perp},\mu)$ the peak location is inconsistent, as the BAO feature for spectroscopic redshifts does not follow the straight $s_{\perp}$ bins.

For Fig. \ref{fig:toymodel_photo} we created the same plots representing photometric redshifts with a redshift uncertainty of $\sigma_{0}=0.02$. The location of the BAO peak becomes inconsistent between different $\mu$ bins in $\xi(s,\mu)$, while it stays stable for the $\xi(s_{\perp},\mu)$ binning scheme. However, a closer look reveals that the center of the peak has shifted to lower values by a few percent. This is because the scatter introduced by the large uncertainty of photometric redshifts mixed parts of the BAO signal originally belonging to different $\mu$ bins together. As the BAO appears as a circular structure that gets washed out along the LOS, the bulk of the remaining feature appears at slightly lower values perpendicular to the LOS. 

\label{sec:shift}
\begin{figure}
    \centering
        \includegraphics[width=0.47\textwidth]{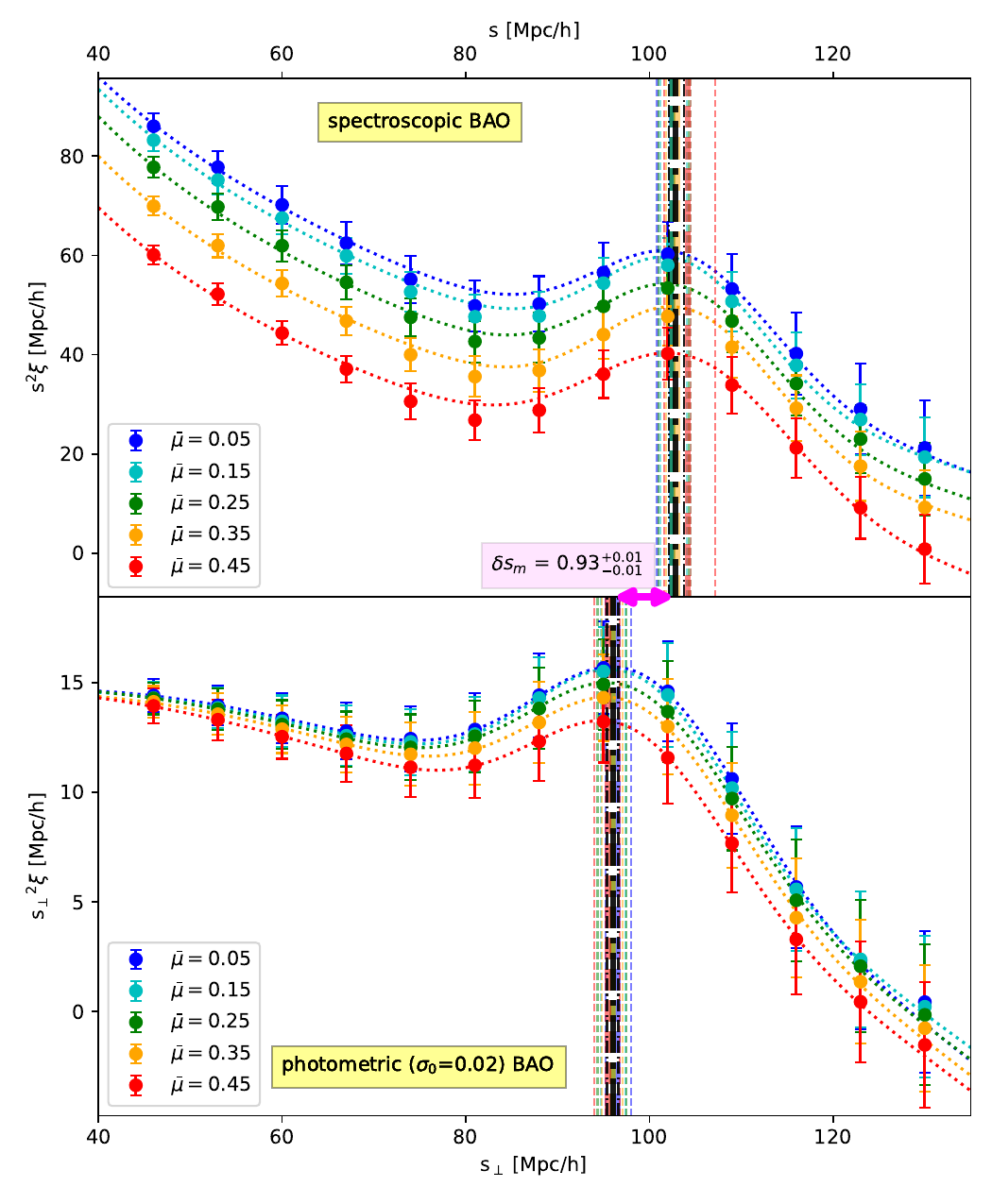}  
        \caption{Comparison of the BAO peak locations in spectroscopic (upper panel) and photometric (lower panel) redshift data. The vertical lines in various colors correspond to the location of the BAO peak found by fitting a six-parameter model to various $\mu$ bins. The vertical black line is the average and its error bar for all $\mu$ bins combined. There is a systematic shift that has to be accounted for by either a correction factor or dedicated templates for photometric redshifts.}
        \label{fig:compare_spec_photo}   
\end{figure} 

Because the toy model, albeit an excellent tool for illustrating the processes that shift the BAO peak, is a highly simplified approach that does not take the full picture of large-scale clustering into account, we needed to repeat the measurements of the BAO peak shift with data from proper numerical simulations. To this end, we carried out the basic tests using the {\tt G2P} simulations produced by our group. We applied the photometric redshift painting (see Sect. \ref{sec:photozpaint}) and calculated the anisotropic correlation function for both the photometric redshifts and spectroscopic redshifts. Afterward, we fit the empirical six-parameter model (see \citealt{Sanchez_empirical,Sanchez_empirical2,Sridhar:2020}) to the mean of the correlation functions of the {\tt G2P} simulations. By comparing the peak locations, which is one of the free parameters, obtained by these fits, we find that the center of the BAO peak is systematically shifted to lower values for the photometric redshifts, as shown in Fig. \ref{fig:compare_spec_photo}. The shift is slightly larger than what we see in the toy model. We obtained the shift by taking the ratio between the photometric BAO peak locations of all $\mu$ bins and the spectroscopic BAO peak location of all $\mu$ bin. This is possible because the location of the peak is consistent in both datasets when using the respective binning schemes and it allows us to measure the value of the shift parameter with high precision. We find a shift parameter $\delta s_{m}$ of $0.93 \pm 0.01$ for the mean of the {\tt G2P} simulations. We repeated the same method with the cut-sky {\tt AbacusSummit} simulations and obtained values consistent with our simulations. 

This shift has to be accounted for in all methods that aim to measure cosmology using the location of the BAO peak. The empirical fits yield a notably different BAO scale for data based on photometric redshifts than for the same data based on spectroscopic redshifts. In the case of template fitting, it is also important to design the templates used specifically for photometric redshift data and not to use ones designed for spectroscopic data. Hence, we created special templates using simulations with photometric redshift uncertainty as shown in Sect. \ref{sec:template_fits}. While testing different levels of photometric redshift uncertainty, we found that the magnitude of the shift does not depend on the exact value of the photometric redshift uncertainty $\sigma_{0}$, given that it is sufficiently large ($\sigma_{0} > \sim 0.01$). This corresponds to the scale found by \citet{Chan:2022} at which the $\xi(s_{\perp},\mu)$ binning scheme starts to yield a consistent BAO peak location across all $\mu$ bins.

\section{Results}
\label{sec:results}
\subsection{Correlation function measurements}
\label{sec:corrfunc_results}
\begin{figure*}
    \centering
        \includegraphics[width=0.95\textwidth]{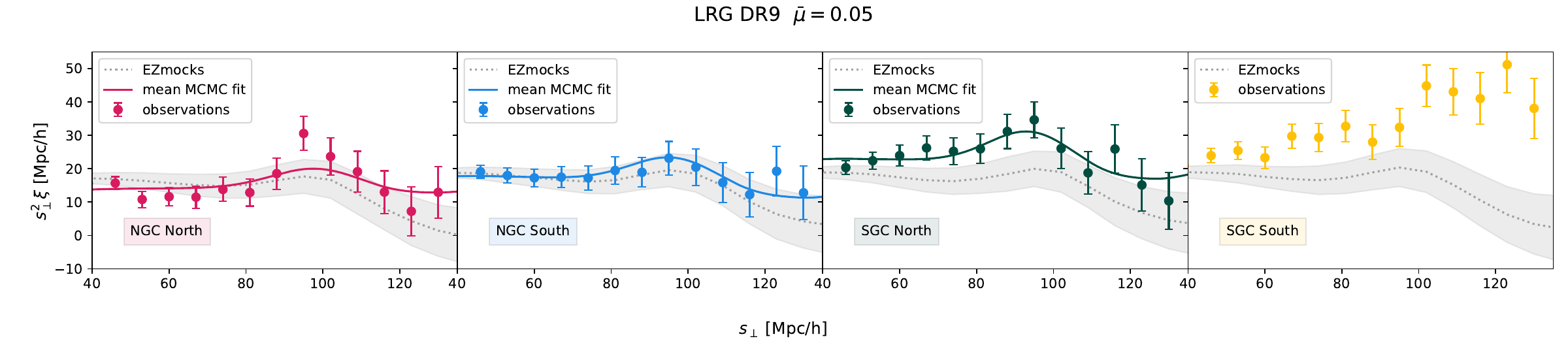}  
        \caption{Correlation functions of the $\bar{\mu}=0.05$ bin for the four subsamples compared to the {\tt EZmocks} correlation functions. }
        \label{fig:results_mu0}   
\end{figure*} 

We analyzed the anisotropic correlation functions for the four subsamples independently and compared them to the predictions obtained from the 1000 {\tt EZmocks} cut to the same geometry. In Fig. \ref{fig:results_mu0} we compare the observed correlation function to the expected one from the {\tt EZmocks} for the first $\mu$ bin. For most subsamples, there is a reasonable agreement between the expected correlation function and the observed ones. Only the observations in SGC South strongly deviate from the expectations. Results are similar for the other $\mu$ bins, which are shown in Figures \ref{fig:results_mu1} to \ref{fig:results_mu4}. 

\subsection{Template fits}
\begin{table}
\begin{center}
\caption{Results of the template fits.} 
\label{tab:template_fit_results} 
\begin{tabular}{rcc|cc}
&Subsample & $\mu_{\textrm{max}}$ & $\alpha_{\perp}$ & $\chi^{2} / DOF $\\
\hline
 &NGC North & 0.1 & $1.013^{+0.024}_{-0.024}$ & 19.41 / 10 \\ 
 &NGC North & 0.2 & $1.021^{+0.019}_{-0.019}$ & 34.41 / 21 \\ 
*&NGC North & 0.3 & $1.012^{+0.020}_{-0.020}$ & 34.1 / 32 \\ 
 &NGC North & 0.4 & $1.018^{+0.023}_{-0.023}$ & 62.09 / 43 \\ 
 &NGC North & 0.5 & $1.032^{+0.030}_{-0.029}$ & 74.79 / 54 \\ 
\hline
 &NGC South & 0.1 & $1.033^{+0.039}_{-0.043}$ & 5.11 / 10 \\ 
 &NGC South & 0.2 & $0.924^{+0.052}_{-0.046}$ & 12.39 / 21 \\ 
*&NGC South & 0.3 & $0.968^{+0.029}_{-0.032}$ & 18.1 / 32 \\ 
 &NGC South & 0.4 & $0.947^{+0.036}_{-0.040}$ & 117.76 / 43 \\ 
 &NGC South & 0.5 & $0.954^{+0.045}_{-0.057}$ & 52.41 / 54 \\ 
\hline
 &SGC North & 0.1 & $1.057^{+0.023}_{-0.024}$ & 10.62 / 10 \\ 
 &SGC North & 0.2 & $1.053^{+0.022}_{-0.022}$ & 16.57 / 21 \\ 
*&SGC North & 0.3 & $1.056^{+0.019}_{-0.020}$ & 25.03 / 32 \\ 
 &SGC North & 0.4 & $1.034^{+0.026}_{-0.027}$ & 131.14 / 43 \\ 
 &SGC North & 0.5 & $1.041^{+0.047}_{-0.050}$ & 81.55 / 54 \\ 
\end{tabular}
\tablefoot{The template fits are combined over multiple $\mu$ bins for the three subsamples with reasonable correlation function measurements. We highlight $\mu_{\textrm{max}}=0.3$ with an asterisk as it provides the best reduced $\chi^{2}$ and will be consequently used in our final analysis.}
\end{center}
\end{table}

We applied the template fits to the three subsamples (NGC North, NGC South, and SGC North) for which we obtained reasonable correlation function measurements. We excluded the fourth subsample (SGC South) from this part of our analysis because the shape of the correlation function is heavily distorted and we would not be able to trust the fits. The results of the fits for the three more reliable subsamples are provided in Table \ref{tab:template_fit_results}. The $\alpha_{\perp}$-parameter measurements were obtained using a joint fit across multiple $\mu$ bins (see Sect. \ref{sec:jointfits}) starting from zero to $\mu_{\textrm{max}}$ as the upper limit of the range $\mu$ considered for the fit. Based on the reduced $\chi^{2}$, we determined that a $\mu_{\textrm{max}}$ of 0.3 is the optimal choice to gain the most reliable results from our data. 

\section{Discussion}
\label{sec:discussion}

\subsection{Comparison of different footprints}
\label{sec:footprintcompare}
As we divided the DR9 footprints into four different areas (see Fig. \ref{fig:skymap}), we compared the peculiarities of the measured correlation functions within them intending to better understand the systematic differences between these footprints. 

The data from NGC North shows a notably exaggerated BAO peak. This is surprising as it is the region with very marginally the lowest quality imaging data and the largest uncertainty on the photometric redshifts. However, aside from the strong BAO peak, the rest of the correlation function for this footprint aligns well with the expected shape from {\tt EZmocks} and the {\tt AbacusSummit} simulations. Such a sharp BAO peak is not seen in any of the other footprints and the fact the data skews toward a strong BAO in all $\mu$ bins indicates that it is more than just sample variance. The correlation function of NGC South pretty much aligns perfectly with one from the {\tt EZmocks}. However, the BAO peak is washed out in some of the $\mu$ bins, most likely due to unfavorable cosmic variance. The shape of the correlation function in SGC North is slightly above the expected correlation function from the {\tt EZmocks}, but the peak is recovered well. The data in the SGC South area is problematic. The upturn in the correlation function at larger separations is indicative of a mask issue, which becomes especially bad below a declination of $-30^\circ$. This implies that the removal of stars and artifacts from foreground sources was not done well for this region, which also lies outside the planned spectroscopic footprint of DESI. We find a similar, albeit less extreme,  behavior for the correlation functions in all footprints (see Figure \ref{fig:results_official_mu0} 
and compare it to Fig. \ref{fig:results_mu0}), when the official LRG target selection \citep{DESI_target_LRG} was applied instead of our slightly modified version from Sect. \ref{sec:sampleselection}. We do not think that this difference will seriously affect the spectroscopic DESI survey, as the spectroscopic redshift data will enable a better final cleaning of the sample than just the photometric redshifts used in our analysis. 

\subsection{Comparison of different $\mu$ bins}
\label{sec:mucompare}
When comparing the correlation functions measured in the different $\mu$ bins for each sample (see Figs. \ref{fig:results_mu0} and \ref{fig:results_mu1} to \ref{fig:results_mu4}
), we found that the BAO feature becomes increasingly more difficult to detect with increasing $\mu$, which is expected and consistent with previous works \citep{Srivatsan:2019}. Using the results presented in Table \ref{tab:template_fit_results}, we found that constraints on $\alpha_{\perp}$ start to become weaker after including the fourth and fifth $\mu$ bins. Hence, we recommend using the $\alpha_{\perp}$ values with a $\mu_{\textrm{max}}$ of 0.3 and disregarding the values that were obtained using data from higher $\mu$ bins as well. We also found the same cutoff point in $\mu_{\textrm{max}}$ when using the {\tt AbacusSummit} cut-sky simulations. With this in mind, we took a closer look at the results of the first three $mu$ bins for the different footprints. In the case of NGC North, the results for the $\alpha_{\perp}$ values stay consistent at 1.012 with error bars that are consistent with the fiducial cosmology within 1$\sigma$. The template fit yields values are compatible (within 1 to 2$\sigma$) with the fiducial cosmology for NGC South for all $\mu$ bins, albeit with larger uncertainties than the other footprints. For the first three $\mu$ bins the results for $\alpha_{\perp}$ for SGC North, we find a moderate tension with the fiducial cosmology of 2 to 3$\sigma$. Overall, we see that our results tend to be compatible with our fiducial cosmology used to create templates, the Planck 2015 cosmology \citep{Planck:2015}. 
 
\subsection{Comparison to other photometric and spectroscopic BAO measurements}
\label{sec:specvsphoto}
\begin{figure}
    \centering
        \includegraphics[width=0.48\textwidth]{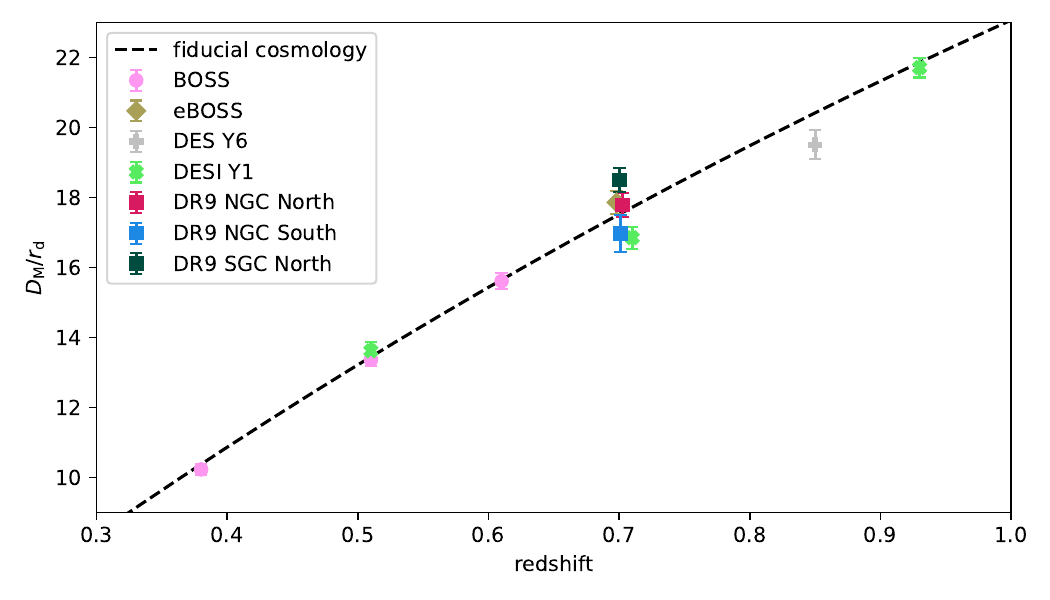}  
        \caption{Comparison of various recent BAO measurements of the angular distance scale within the intermediate redshift range. The dashed black line shows  the fiducial cosmology of our paper, pink circles the BOSS data \citep{Alam_2017}, the olive diamond the eBOSS data \citep{Bautista:2021}, the gray plus-shaped point the DES Y6 data \citep{DES_Y6_BAO}, lime-green crosses the DESI Y1 data \citep{DESI_2024_3}, and squares the data presented in this paper for NGC North (red), NGC South (blue), and SGC North (dark green).}
        \label{fig:cosmology}   
\end{figure} 

\begin{table}
\begin{center}
\caption{Comparison of various recent BAO measurements of the angular distance scale within the intermediate redshift range. } 
\label{tab:cosmology} 
\begin{tabular}{c|cc}
Subsample & $z_{\textrm{eff}}$ & $D_{\textrm{M}}/r_{\textrm{d}}$ \\
\hline
DR9 NGC North &  0.703  & $ 17.79  \pm  0.35  $ \\
DR9 NGC South &  0.701  & $ 16.97  \pm  0.54  $ \\
DR9 SGC North &  0.700  & $ 18.50  \pm  0.34  $ \\
\hline
BOSS LOWZ+CMASS &  0.38  & $ 10.23  \pm  0.15  $ \\
BOSS LOWZ+CMASS &  0.51  & $ 13.37  \pm  0.18  $ \\
BOSS LOWZ+CMASS &  0.61  & $ 15.61  \pm  0.22  $ \\
eBOSS LRG &  0.698  & $ 17.86  \pm  0.33  $ \\
DES Y6 &  0.85  & $ 19.51  \pm  0.41  $ \\
DESI Y1 &  0.51  & $ 13.62  \pm  0.25  $ \\
DESI Y1 &  0.71  & $ 16.85  \pm  0.32  $ \\
DESI Y1 &  0.93  & $ 21.71  \pm  0.28  $ \\
DESI Y1 &  1.32  & $ 27.79  \pm  0.69  $ \\
\end{tabular}
\end{center}
\end{table}
The main advantages of photometric surveys and photometric redshifts are that they are an efficient way to cover a large area quickly, while spectroscopic redshift surveys can reach much higher levels of precision at the cost of significantly more observing time and the requirement of preexisting imaging surveys for targeting. However, the differences between BAO measurements obtained from spectroscopic redshifts and ones obtained from photometric redshifts are more than just data quality. There are various systematic biases, and properly addressing them requires the use of different methods. \citet{Chan:2022} showed that for redshift uncertainties that are typical for photometric redshifts from broadband imaging surveys, such as the DESI Legacy Imaging Survey DR9, the binning scheme used to obtain the anisotropic correlation function from spectroscopic survey data yields inconsistent BAO peak locations between the different $\mu$ bins. The suggested alternative binning strategy, calculating $\xi(s_{\perp},\mu)$ instead of the usual $\xi(s,\mu),$ yields consistent BAO peak locations, and we can hereby confirm the results of \citet{Chan:2022} in this regard. However, on top of this issue, we discovered that the BAO peak location is systematically shifted in the case of data with typical photometric redshift uncertainty when compared to data with spectroscopic redshift quality as discussed in great detail in Sect. \ref{sec:peak_shift}. This shift needs to be considered when comparing observations to theoretical predictions or templates from simulations. Both discussed effects were not considered in previous works on DESI Legacy Imaging Survey data such as in \citet{Sridhar:2020}. The samples that we considered are listed in Table \ref{tab:cosmology} and illustrated in Fig. \ref{fig:cosmology}. The DES Y3 and Y6 footprint \citep{DES_3Y_photo,DES_Y6_BAO_sample} is located in the SGC and is within our SGC North and SGC South footprints. \citet{DES_Y3_BAO} and \citet{DES_Y6_BAO} found, using a very similar method, results comparable to ours, but at a higher effective redshift. The combined BOSS low-z and CMASS samples cover a lower redshift range than our sample. However, the eBOSS LRG sample \citep{Ross:2020} has an effective redshift of 0.698 \citep{GilMarin:2020} almost the same as our samples. Although the eBOSS LRG sample extends into all three footprints for which we have measured the BAO scale, it is dominated and the most complete in the areas belonging to NGC North that contains $\sim 40 \%$ of the galaxies and to a lesser extent in SGC North ($\sim 32 \%$ of the galaxies) and also NGC South ($\sim 28 \%$ of the galaxies). In the case of the DESI Y1 data \citep{DESI_2024_3}, most of it ($\sim 53 \%$) can be found in NGC South, which the LRG sample within our redshift range shows a similar low value in our photometric data as it does for the spectroscopic Y1 data. Less of the DESI Y1 data are distributed in SGC North ($\sim 31 \%$ ) and  NGC North  ($\sim 17 \%$).  Therefore, our BAO measurements using photometric redshifts support the assumption that the apparent difference between final eBOSS and DESI Y1 BAO measurements is caused by cosmic variance due to the footprints covered by both surveys. 

\section{Summary and conclusions}
\label{sec:summary}
We carried out measurements of the correlation functions for four distinct footprints within the DESI Legacy Imaging Survey DR9 and discussed the differences between them. To this end, we used the DESI LRG target selection with additional masks to better clean the sample, and we further limited our sample to a photometric redshift range between 0.6 and 0.8. While we find that SGC South suffers from serious masking issues, we managed to obtain BAO measurements in all other footprints. We also discussed the systematic difference between BAO measurements that rely on photometric redshifts with their relatively large uncertainties and high-quality spectroscopic redshifts. In addition to the known issues in measuring the correlation function and the necessary modifications, we also find that the location of BAO peaks is systematically shifted in the case of photometric redshifts. We considered that shift when constructing fitting templates from simulations. 

The key results of this paper are that our most robust measurements ($\mu_{\textrm{max}}$=0.3) provide a possible explanation for the apparent tension between the eBOSS LRG \citep{Bautista:2021} and DESI Y1 LRG samples \citep{DESI_2024_3}   within the same redshift range. As we have measurements for the comoving angular distance in the three distinct footprints that make up the survey area for the final DESI release and therefore also contain the eBOSS footprint, we can study the variance of the BAO measurements at the critical redshift range for the tension within these areas. The value we obtain in NGC South, which covers the area in which most of the spectroscopic DESI Y1 data can be found, yields a value very similar to that in DESI Y1, while the areas in which most of the eBOSS LRG sample is located yield values comparable to the eBOSS LRG values (see Fig. \ref{fig:cosmology}). Given the nature of measuring BAOs with photometric redshifts, we can only probe the comoving angular distance and not the Hubble rate. However, given the strong agreement of our data with eBOSS and DESI in their respective footprints, we argue that the apparent tension is mainly caused by sample variance across the sky. Therefore, we expect that for the final DESI release, which will cover the three footprints discussed in this paper, the BAO measurements will yield a value compatible with the average of all of them.

\section*{Data availability}
All data points shown in the figures are available in a machine-readable form at \href{https://doi.org/10.5281/zenodo.14733076}{https://doi.org/10.5281/zenodo.14733076}.
\begin{acknowledgements}

This work was supported by the National Research Foundation of Korea (NRF) grant funded by the Korea government (MIST), Grant No. RS-2021-NR058702. YZ acknowledges the support from the National Natural Science Foundation of China (NFSC) through grant 12203107, the Guangdong Basic and Applied Basic Research Foundation with No.2019A1515111098, and the science research grants from the China Manned Space Project with NO.CMS-CSST-2021-A02. Data analysis was performed using the high-performance computing cluster {\tt Seondeok} at the Korea Astronomy and Space Science Institute. \\

This material is based upon work supported by the U.S. Department of Energy (DOE), Office of Science, Office of High-Energy Physics, under Contract No. DE–AC02–05CH11231, and by the National Energy Research Scientific Computing Center, a DOE Office of Science User Facility under the same contract. Additional support for DESI was provided by the U.S. National Science Foundation (NSF), Division of Astronomical Sciences under Contract No. AST-0950945 to the NSF’s National Optical-Infrared Astronomy Research Laboratory; the Science and Technology Facilities Council of the United Kingdom; the Gordon and Betty Moore Foundation; the Heising-Simons Foundation; the French Alternative Energies and Atomic Energy Commission (CEA); the National Council of Humanities, Science and Technology of Mexico (CONAHCYT); the Ministry of Science, Innovation and Universities of Spain (MICIU/AEI/10.13039/501100011033), and by the DESI Member Institutions: \url{https://www.desi.lbl.gov/collaborating-institutions}. Any opinions, findings, and conclusions or recommendations expressed in this material are those of the author(s) and do not necessarily reflect the views of the U. S. National Science Foundation, the U. S. Department of Energy, or any of the listed funding agencies.\\

The DESI Legacy Imaging Surveys consist of three individual and complementary projects: the Dark Energy Camera Legacy Survey (DECaLS), the Beijing-Arizona Sky Survey (BASS), and the Mayall z-band Legacy Survey (MzLS). DECaLS, BASS and MzLS together include data obtained, respectively, at the Blanco telescope, Cerro Tololo Inter-American Observatory, NSF’s NOIRLab; the Bok telescope, Steward Observatory, University of Arizona; and the Mayall telescope, Kitt Peak National Observatory, NOIRLab. NOIRLab is operated by the Association of Universities for Research in Astronomy (AURA) under a cooperative agreement with the National Science Foundation. Pipeline processing and analyses of the data were supported by NOIRLab and the Lawrence Berkeley National Laboratory. Legacy Surveys also uses data products from the Near-Earth Object Wide-field Infrared Survey Explorer (NEOWISE), a project of the Jet Propulsion Laboratory/California Institute of Technology, funded by the National Aeronautics and Space Administration. Legacy Surveys was supported by: the Director, Office of Science, Office of High Energy Physics of the U.S. Department of Energy; the National Energy Research Scientific Computing Center, a DOE Office of Science User Facility; the U.S. National Science Foundation, Division of Astronomical Sciences; the National Astronomical Observatories of China, the Chinese Academy of Sciences and the Chinese National Natural Science Foundation. LBNL is managed by the Regents of the University of California under contract to the U.S. Department of Energy. The complete acknowledgments can be found at \url{https://www.legacysurvey.org/}.\\

Any opinions, findings, and conclusions or recommendations expressed in this material are those of the author(s) and do not necessarily reflect the views of the U. S. National Science Foundation, the U. S. Department of Energy, or any of the listed funding agencies.\\

The authors are honored to be permitted to conduct scientific research on Iolkam Du’ag (Kitt Peak), a mountain with particular significance to the Tohono O’odham Nation. \\

This research made use of the following python packages: {\sc astropy}\citep{astropy_2013,astropy_2018,astropy:2022}, {\sc numpy} \citep{numpy}, {\sc corrfunc} \citep{corrfunc,corrfunc2}, {\sc scipy} \citep{scipy}, {\sc emcee} \citep{emcee}, and {\sc matplotlib} \citep{matplotlib}. 

\end{acknowledgements}

%
%
\bibliographystyle{aa}
\bibliography{main} 

\begin{appendix} 
\section{Additional Figures}\label{sec:appendix1}

\begin{figure*}
    \centering
       \includegraphics[width=0.95\textwidth]{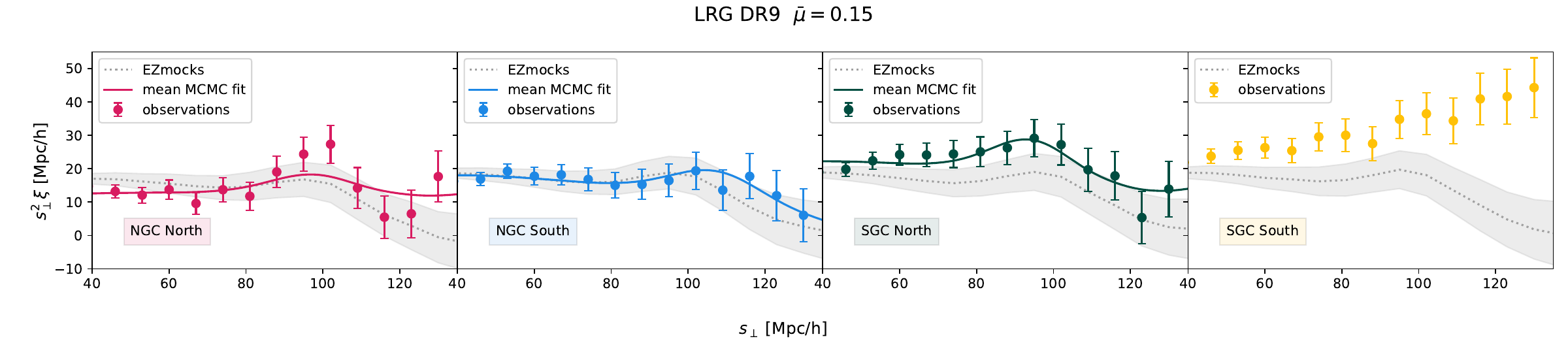}  
       \caption{The correlation functions of the $\bar{\mu}=0.15$ bin for the four sub-samples compared to the EZmock correlation functions. }
       \label{fig:results_mu1}   
\end{figure*}

\begin{figure*}
    \centering
       \includegraphics[width=0.95\textwidth]{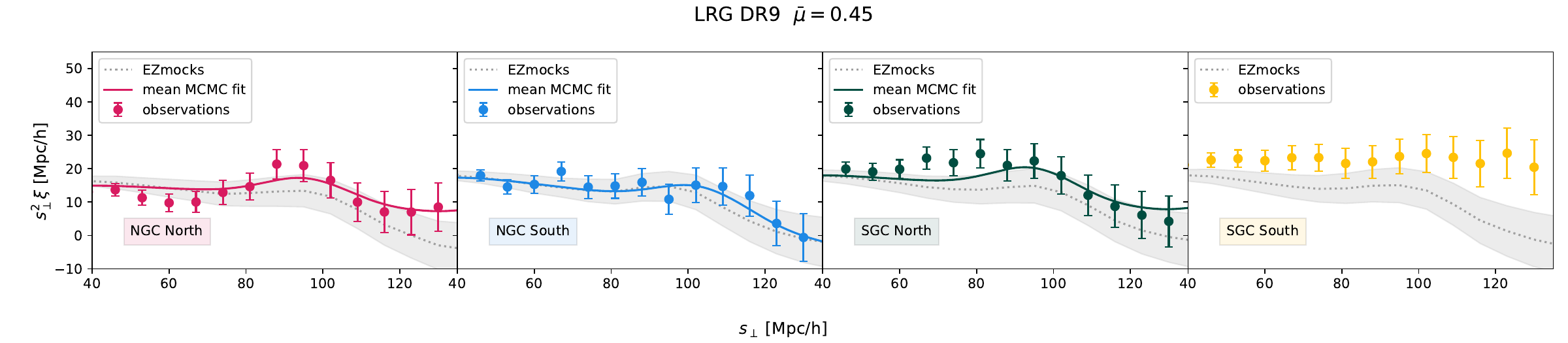}  
       \caption{The correlation functions of the $\bar{\mu}=0.45$ bin for the four sub-samples compared to the EZmock correlation functions. }
       \label{fig:results_mu4}   
\end{figure*}

\begin{figure*}
    \centering
       \includegraphics[width=0.95\textwidth]{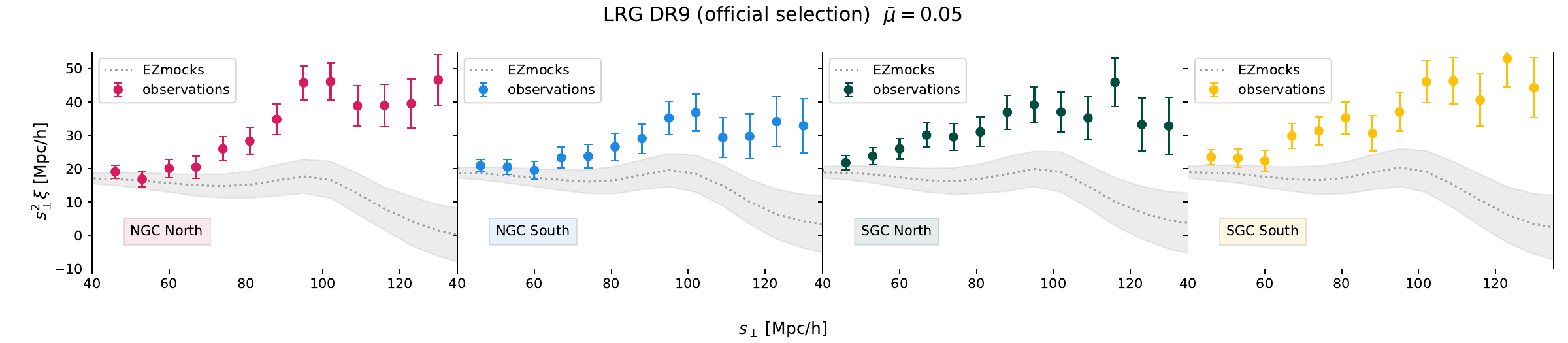}  
       \caption{The correlation functions of the $\bar{\mu}=0.05$ bin for the four sub-samples compared to the EZmock correlation functions, but using the official definition of DESI LRGs instead of our slightly modified version. }
       \label{fig:results_official_mu0}   
\end{figure*}

While in Section \ref{sec:corrfunc_results}, we only provide the plot for the observed correlation function and {\tt EZmocks} of the first $\mu$ bin in Figure \ref{fig:results_mu0}, we put the correlation functions for the remaining $\mu$ bins (Figures \ref{fig:results_mu1} to  to \ref{fig:results_mu4}) in this appendix. Figure \ref{fig:results_official_mu0} aims to compare the correlation function measurements using the official DESI LRG selection to that presented in Figure \ref{fig:results_mu0}.

\begin{figure*}
    \centering
       \includegraphics[width=0.95\textwidth]{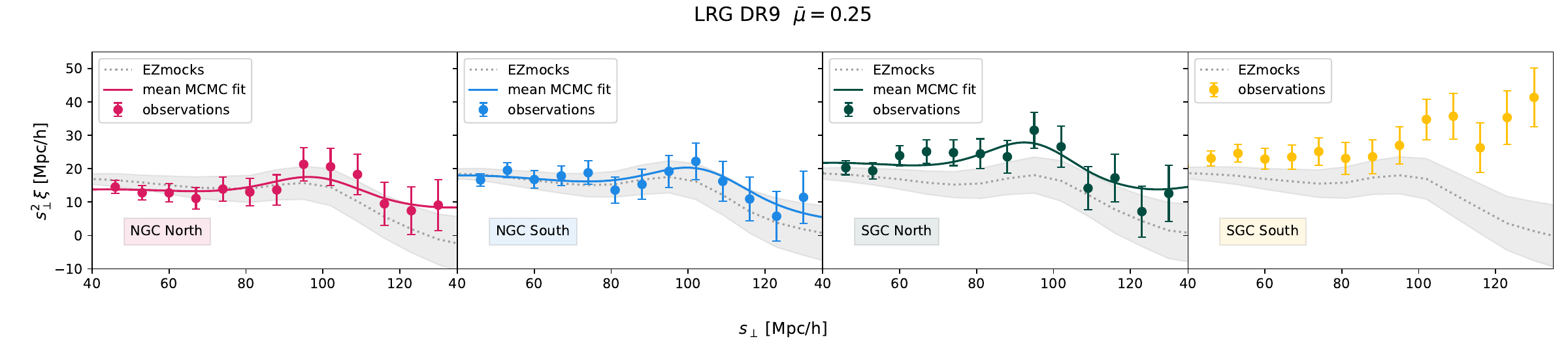}  
       \caption{The correlation functions of the $\bar{\mu}=0.25$ bin for the four sub-samples compared to the EZmock correlation functions. }
       \label{fig:results_mu2}   
\end{figure*} 
\begin{figure*}
    \centering
       \includegraphics[width=0.95\textwidth]{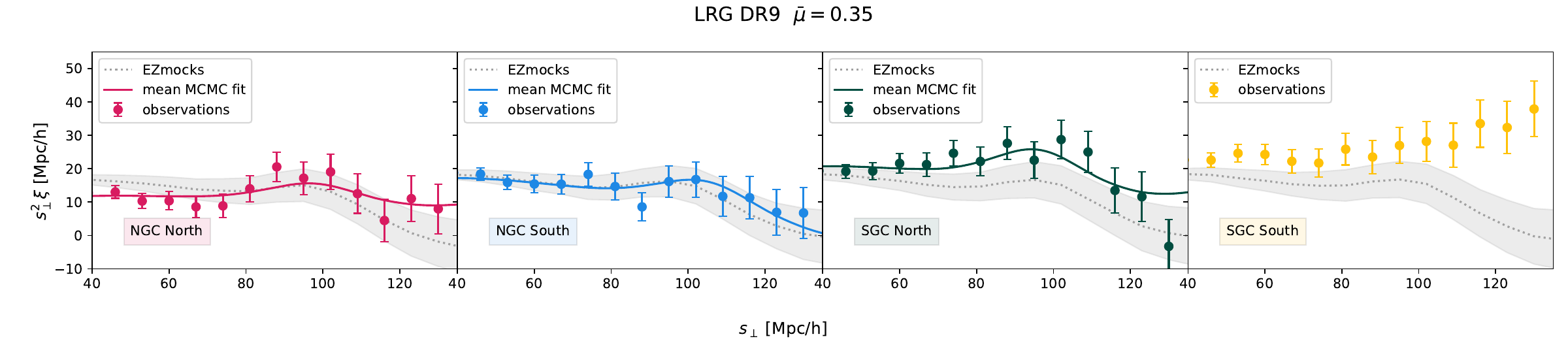}  
       \caption{The correlation functions of the $\bar{\mu}=0.35$ bin for the four sub-samples compared to the EZmock correlation functions. }
       \label{fig:results_mu3}   
\end{figure*} 

\end{appendix}


\end{document}